\documentclass[fleqn,10pt]{wlscirep}
\usepackage[utf8]{inputenc}
\usepackage[T1]{fontenc}

\newcommand{\STO}{SrTiO$_3$}
\newcommand{\LAO}{LaAlO$_3$}

\makeatother
\title{Concomitant appearance of conductivity and superconductivity in (111)  LaAlO$_3$/SrTiO$_3$ interface with metal capping}

\author[1]{R. S. Bisht}
\author[1]{M. Mograbi}
\author[1]{P. K. Rout}
\author[1]{G. Tuvia}
\author[1,*]{Y. Dagan}
\author[2,3]{Hyeok Yoon}
\author[2,3]{A. G. Swartz}
\author[2,3]{H. Y. Hwang}
\author[4]{L. L. Li}
\author[4]{R. Pentcheva}

\affil[1]{School of Physics and Astronomy, Tel-Aviv University, Tel Aviv, 6997801, Israel}
\affil[2]{Department of Applied Physics, Geballe Laboratory for Advanced Materials, Stanford University, 476 Lomita Mall, Stanford, CA 94305, USA}
\affil[3]{Stanford Institute for Materials and Energy Sciences, SLAC National Accelerator Laboratory, Menlo Park, California 94025, USA}
\affil[4]{Department of Physics and Center for Nanointegration Duisburg-Essen (CENIDE), University of Duisburg-Essen, Lotharstr. 1, D-47057 Duisburg, Germany}

\affil[*]{yodagan@tauex.tau.ac.il}

\begin{abstract}
 In polar-oxide interfaces, a certain number of monolayers (ML) is needed for conductivity to appear. This threshold for conductivity is explained by accumulating sufficient electric potential to initiate charge transfer to the interface. Here we study experimentally and theoretically the (111) \STO/\LAO~ interface where a critical thickness, $t_c$, of nine epitaxial \LAO~ ML is required to turn the interface from insulating to conducting and even superconducting. We show that $t_c$ decreases to 3ML when depositing a cobalt over-layer (capping) and 6ML for platinum capping. The latter result contrasts with the (100) interface, where platinum capping increases $t_c$ beyond the bare interface. The observed threshold for conductivity for the bare and the metal-capped interfaces is confirmed by our density functional theory calculations. Interestingly, for (111) \STO/\LAO/Metal interfaces, conductivity appears concomitantly with superconductivity in contrast with the (100) \STO/\LAO/Metal interfaces where $t_c$ is smaller than the critical thickness for superconductivity. We attribute this dissimilarity to the different orbital polarization of $e_g^\prime$ for the (111) versus $d_{xy}$ for the (001) interface.
\end{abstract}
\begin{document}

\flushbottom
\maketitle

\thispagestyle{empty}

\section*{Introduction}

The interface between \LAO~ and \STO~ exhibits  two-dimensional conductivity \cite{hwang2004}, superconductivity \cite{Reyren1196}, magnetism \cite{brinkman, sachs2010anomalous, bert2011,RonMagnetism2014,KleinAMR,lee2013titanium}, metal-insulator transition \cite{Thiel1942}, tunable Rashba spin-orbit interaction \cite{BenShalom2010,caviglia2010tunable}, quantum hall states \cite{xie2014quantum,trier2016quantization}, and one-dimensional conductivity \cite{ron2014one,briggeman2020pascal}. While  the (100) \STO/\LAO~ interface received significant scientific attention, the (111) interface remains less explored.
\par
Along the [111] direction in \STO\ Ti forms triangular layers \cite{herranz2012, khanna2019symmetry}. Due to the trigonal symmetry the  degeneracy of the t$_{2g}$ manifold is lifted and they are further split into a$_{1g}$ and e$'_g$ orbitals. The six-fold symmetry of the titanium layer is reflected in the transport properties \cite{YoramAMR111}. The symmetry is further reduced due to the structural transition of \STO~and the interface is predicted to host exotic superconductivity \cite{hecker2018vestigial}, and topological states \cite{111DFT1}.
\par
Four monolayers of \LAO~ are needed for the formation of a two-dimensional electron system (2DES) at the (100) \STO/\LAO~ interface \cite{Thiel1942}. For the (111) interface, the critical thickness for conductivity is nine monolayers (ML) \cite{herranz2012}. The (111) 2DES also exhibits superconductivity \cite{caviglia111sc}, with a link between superconductivity and spin-orbit interaction \cite{Rout111}. Notably, upon carrier depletion with negative gate voltage, superconductivity transitions into a Bose-insulating state \cite{morgbiSIT}. This behavior contrasts with the (100) interface where a weaker insulating state is observed for negative gate biases \cite{caviglia2008,chen2018carrier}. 
\par
For spin injection and low voltage transistor applications, the barrier produced by the minimal four monolayers of \LAO~ required for conductivity at the bare (100) interface or by the nine monolayers at the (111) interface is relatively strong. Recent first principles calculations \cite{MetalcapDFT}, and experimental studies of various metal capping on (100) interfaces (\STO/\LAO/Metal)  \cite{MetalCapXLD,MetalcapElectrochemistry} show that the critical thickness for the onset of conductivity, $t_c$, can be reduced relative to the bare interface and that $t_c$ increases with the work function. 
\par
Here we study the problem of critical thickness for conductivity for (111) interfaces both experimentally and theoretically. We have also expanded our experimental research to the superconducting properties of both (100) and (111) \STO/\LAO/metal interfaces. We find that upon capping the (111) \STO/\LAO~ interface with cobalt (Co) and platinum (Pt), $t_c$ is reduced from $t_c$(bare)=9 ML to   $t_c$(Co)=3 ML and $t_c$(Pt)=6 ML. Furthermore, once the (111) interface becomes conducting, it also becomes superconducting at low temperatures. This contrasts with the (100) interface where $t_c$(Pt)>$t_c$(bare) and conductivity appears at $t_c$ without superconductivity.
\par
Concomitant density functional theory calculations with a Hubbard $U$ term (DFT+$U$) confirm the reduction of the critical thickness upon metal capping and indicate that conductivity at the (111) interface arises due to bands with $e_g^\prime$ orbital polarization. We conjecture that these bands are also responsible for superconductivity. This is in contrast with the (100) interface where the $d_{xy}$ and the $d_{yz}, d_{xz}$ bands are split due to their different effective masses along the direction of the confining potential ($z$-direction) \cite{santander2011}. However, atomic spin-orbit interaction mixes them together \cite{joshua2012}. This new band-structure becomes non-rigid due to a Hubbard-type on-site repulsion with a lower-energy, non-superconducting band, and a higher-energy, mobile band, which is responsible to superconductivity \cite{khanna2019symmetry, maniv100}.  

\section*{Results}
\subsection*{Experimental results}
\begin{figure}[h]
 \centering
\includegraphics[width=8.3 cm]{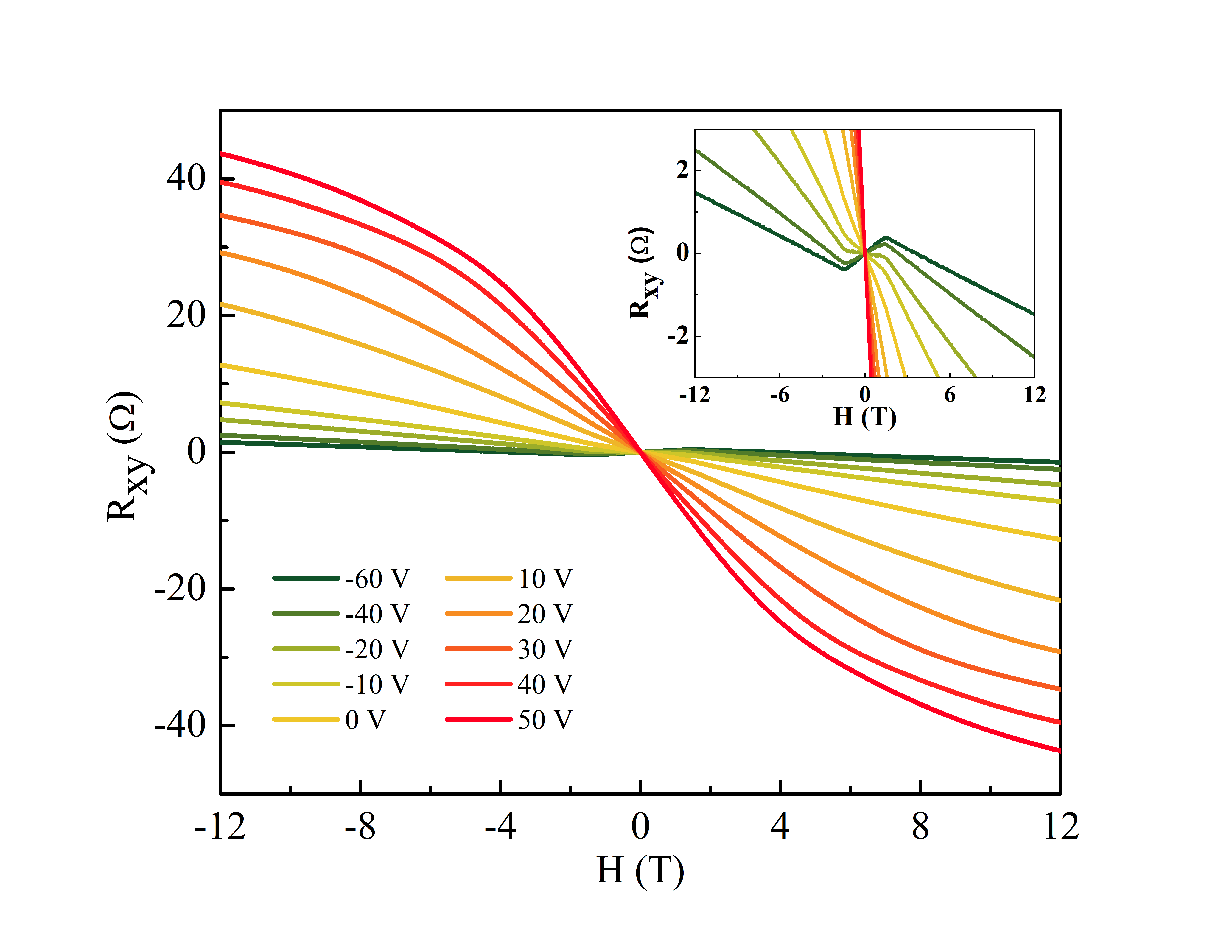}
\caption{Transverse resistance of (111)\STO/\LAO/Co/AlO$_x$ for 4 ML \LAO~  as a function of a perpendicular magnetic field at different gate voltages. The inset focuses on the negative gate voltage regime. The observed anomalous-Hall signal demonstrates the predominance of the Co layer properties in this regime. This is in contrast to the higher carrier density regime where the 2DES dominates.\label{Hall}}
\end{figure}

The transport measurements were performed on the samples with metal capping. In this configuration, the measured resistance is either a parallel combination of the metal cap resistance and the 2DES at the conducting \STO/\LAO~ interface or only the metallic cap in the absence of 2DES. We demonstrate this by measuring the transverse resistance R$_{xy}$, i.e., the Hall signal of the (111) \STO/\LAO/Co/AlO$_x$ interface for 4 \LAO~ ML, as shown in Fig.\ref{Hall}. While for positive gate voltage, the 2DES dominates and exhibits signal resembling the (111) \STO/\LAO~ interface \cite{Rout111} when depleting the 2DES by negative gate voltage, the contribution of Co predominates as manifested in an anomalous-Hall signal, confirming the presence of two parallel channels for conduction. This behavior is similar to (100) with cobalt capping \cite{MetalcapElectrochemistry}.
\par
\begin{figure}[h]
\centering
\includegraphics[width=8.3 cm]{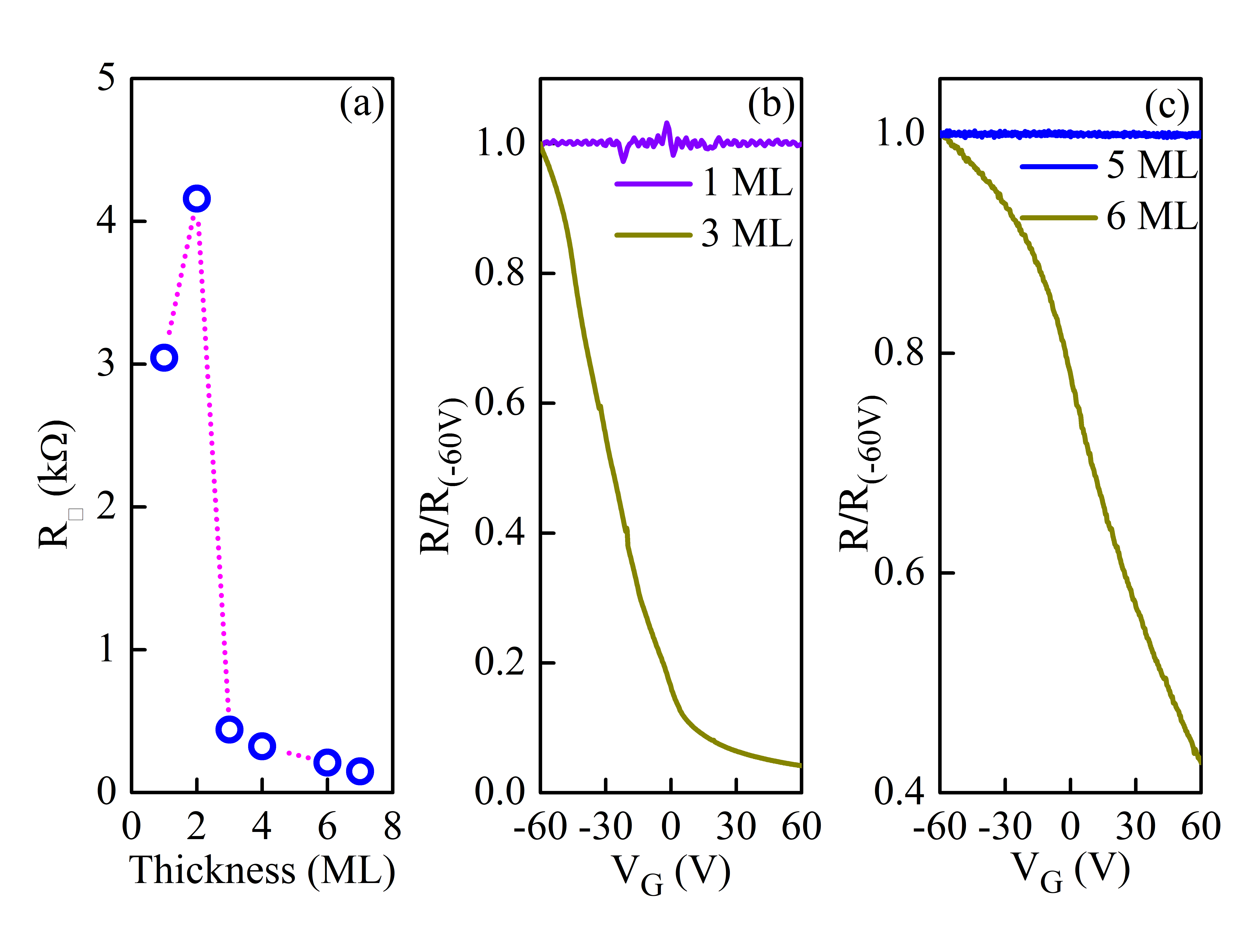}
\caption{(a) The sheet resistance ($R_{\Box}$) of (111) \STO/\LAO/Co/AlO$_x$ at 40 K as a function of \LAO~ thickness, an abrupt drop in the resistance at 3 ML indicates that this is the critical thickness for the formation of 2DES at the \STO/\LAO~interface. Note: The measured samples were not patterned, the geometrical factor for the sheet resistance calculation is within an error bar of $\pm10\%$. (b) The gate dependence of the normalized resistance (R/R$_{(-60V)}$) for (111) \STO/\LAO/Co/AlO$_x$ with \LAO~ thickness of 1 and 3 monolayers.  (c) The gate dependence of normalized resistance (R/R$_{(-60V)}$) for (111) \STO/\LAO/Pt for \LAO~ thickness of 5 and 6 monolayers. \label{Transport-propt}}
\end{figure}

The transport studies conducted on (111) \STO/\LAO/Co/AlO$_x$ interface show a reduction of \LAO~ critical thickness for the onset of 2DES conductivity from 9 ML to 3 ML. In Fig.\ref{Transport-propt} (a) we show the sheet resistance of (111)\STO/\LAO/Co/AlO$_x$ as a function of \LAO~ thickness at 40 K. For \LAO~ thickness below 3 ML, the resistance increases by nearly a factor of five. This indicates that 3 ML is the critical thickness of \LAO~ for the onset of conductivity with Co capping ($t_c$(Co)=3 ML). To verify that a 2DES is formed parallel to the metallic layer, we measured the resistance versus back gate voltage. For a thin metallic layer parallel to a 2DES, one expects gate-dependent resistance due to the dominating contribution of 2DES. On the other hand, in the absence of a 2DES parallel to a metallic layer, we expect the gate dependence of the resistance to be immeasurably small due to the substantial carrier density in the metal. Fig.~\ref{Transport-propt} (b) shows the gate dependence of (111) \STO/\LAO/Co/AlO$_x$ for 1 and 3 \LAO~ ML. For the sample with a single \LAO~ ML, the normalized resistance (R/R$_{(-60V)}$) is flat as a function of gate voltage, suggesting the absence of 2DES at the \STO/\LAO~ interface. For 3 ML \LAO, the data show a significant gate dependence, suggesting the formation of a 2DES at (111)\STO/\LAO~ interface. We conclude that $t_c$ becomes 3 ML upon Co capping for (111) interface.
\par
\par
Previous studies on the metal-capped (100) \STO/\LAO~ interface show that $t_c$ increases with the work function of the metal-capping layer \cite{MetalcapDFT}. To understand the role of the work function in (111) interfaces, we carried out experiments with Pt capping. The work function of platinum is generally higher than that of Co \cite{herbertworkfunction}. Surprisingly, we found that, unlike the (100) interface, Pt capping reduces the critical thickness from $t_c$(bare)=9 ML to $t_c$(Pt)=6 ML. This is demonstrated in Fig.\ref{Transport-propt}(c), where we show the gate dependence of (111)\STO/\LAO/Pt for 5 and 6 \LAO~ ML. The absence of gate dependence for 5ML and the strong gate dependence for 6ML suggests that $t_{c}$(Pt)=6 ML. We interpret the saturation of the resistance at negative gate voltage as a result of depletion of the 2DES and dominance of the metal capping layer (see figure S2 in the supplementary information). In Figure S3 (Supplementary information), we also show the gate dependence of the extracted sheet resistance of 2DES for (111)\STO/\LAO/Co/AlO$_x$ and  (111)\STO/\LAO/Pt for 3 ML and 6 ML of \LAO~respectively.
\par
While suppression of $t_c$ upon Co capping is observed for both the (111) and (100) \STO/\LAO~ interfaces, a reduction of $t_c$ upon Pt capping is observed only for (111) interface, whereas an increase in $t_{c}$ is found for the Pt capped (100) interface \cite{MetalcapElectrochemistry}. 

\begin{figure*}[htbp]
\centering
\includegraphics[width=0.65\textwidth]{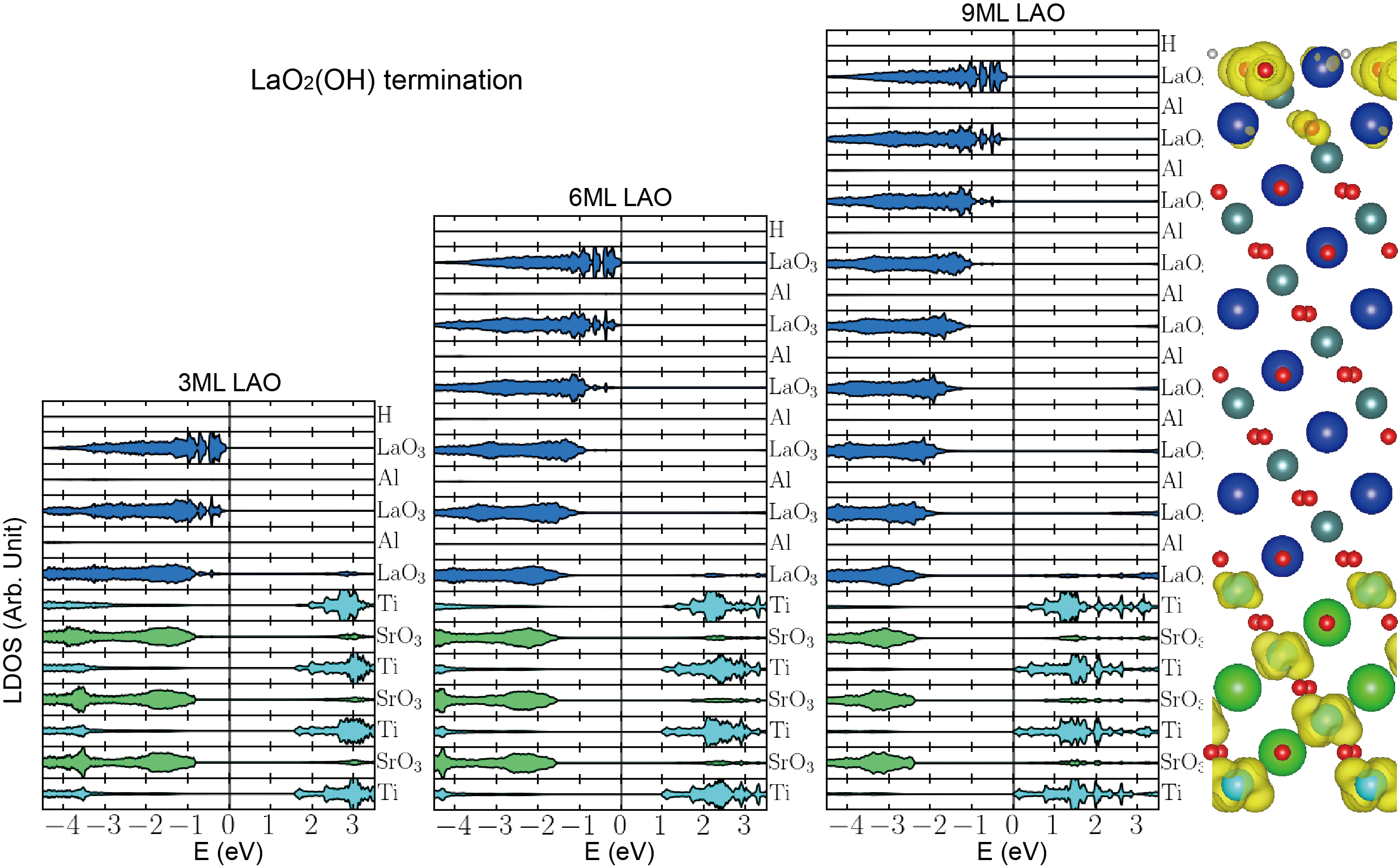}
\caption{Layer-resolved density of states (LDOS) of (111)\STO/\LAO slabs containing 7 ML of STO(111) and $N$ ML of \LAO~ with a LaO$_2$(OH) termination ($N=3$, 6, 9). A side view of half of the simulation cell and the electron density distribution integrated from -0.2 to the Fermi level for 9 ML LAO is shown in the right panel.}
\label{LDOS1}
\end{figure*}
\subsection*{DFT results}
In order to shed light on the origin of the experimentally observed critical thickness for an insulator-to-metal transition of the uncovered \LAO\ films on (111)\STO~ as well as the role of metal capping, we have performed density functional theory calculations with a Hubbard $U$ term. We considered different \LAO~ thicknesses and surface terminations, as well as metal capping using symmetric slabs with up to 130 atoms in the simulation cell. In the (111)-direction, the perovskite structure comprises a stacking of AO$_3$-B layers. Taking into account the formal charges in these compounds, this leads to alternating (SrO$_3$)$^{4-}$ and Ti$^{4+}$ layers in \STO~ versus (LaO$_3$)$^{3-}$ and Al$^{3+}$ layers in \LAO~. In analogy to the (001) interface, one expects a Ti termination of (111)\STO, on top of which \LAO~ is deposited. However, for an integer number of \LAO~ layers, the resulting Al-termination resulted in an opposite internal electric field within \LAO~ than the one known from the (001)-oriented films\cite{pentcheva2009,pentcheva2010,singhbhala2018} with the valence band maximum in the top layer pushed to lower energies. This is, however, fully consistent with the charges of the stacked layers along the [111] direction. Moreover, the system is metallic as the bottom of the Ti $3d$ band touches the Fermi level, independent of the \LAO~ thickness, thus showing no indication for an insulator-to-metal transition with \LAO~ thickness in contrast with the experimental observation. Therefore, this configuration was discarded. We explored next a LaO$_3$ termination: While this generates an internal electric field with the correct sign with an upward shift of the valence band maximum in the \LAO~ layers as they approach the surface and a gradual reduction of the bandgap with increasing \LAO~ thickness, it generates holes in the surface LaO$_3$ layer. As a result, the Ti $3d$ band at the interface is pushed to higher energies, leading to a higher critical thickness than the one observed in the experiment. This indicates that some kind of compensation of the holes in the surface LaO$_3$ layer is necessary.
To simulate this effect, we have tried different concentrations of hydrogen (H) in the surface layer and identified that a LaO$_2$(OH) termination with a single H on one out of three oxygens as displayed in Fig. \ref{LDOS1} provides the correct charge to suppress the surface holes and to induce the insulator-metal transition for $t_c$(bare)=9 ML \LAO~ on (111)\STO, in agreement with the experiment. We note that the compensation of the surface charge can also be achieved in other ways than hydroxylation - e.g., by creating half an oxygen vacancy per  $(1\times 1)$ surface cell. Surface vacancies were proposed as a possible origin of the metal-to-insulator transition for the (001) oriented case \cite{Zunger} and recently explored for the (111)-oriented interfaces~\cite{min2021cooperative}. Since the modeling of the correct concentration of surface vacancies requires a doubling of the already large unit cell, we continue the analysis with the LaO$_3$(OH) termination, which leads to the qualitatively correct critical thickness. As displayed in Fig. \ref{LDOS1} the band gap of STO(111)/$N$LAO gradually decreases from 1.7 eV for 3 ML \LAO~ to 0.9 eV for 6 ML \LAO~ and is finally quenched for 9 ML \LAO~ consistent with the experimental observation.
\begin{figure*}[htbp]
\centering
\includegraphics[width=0.65\textwidth]{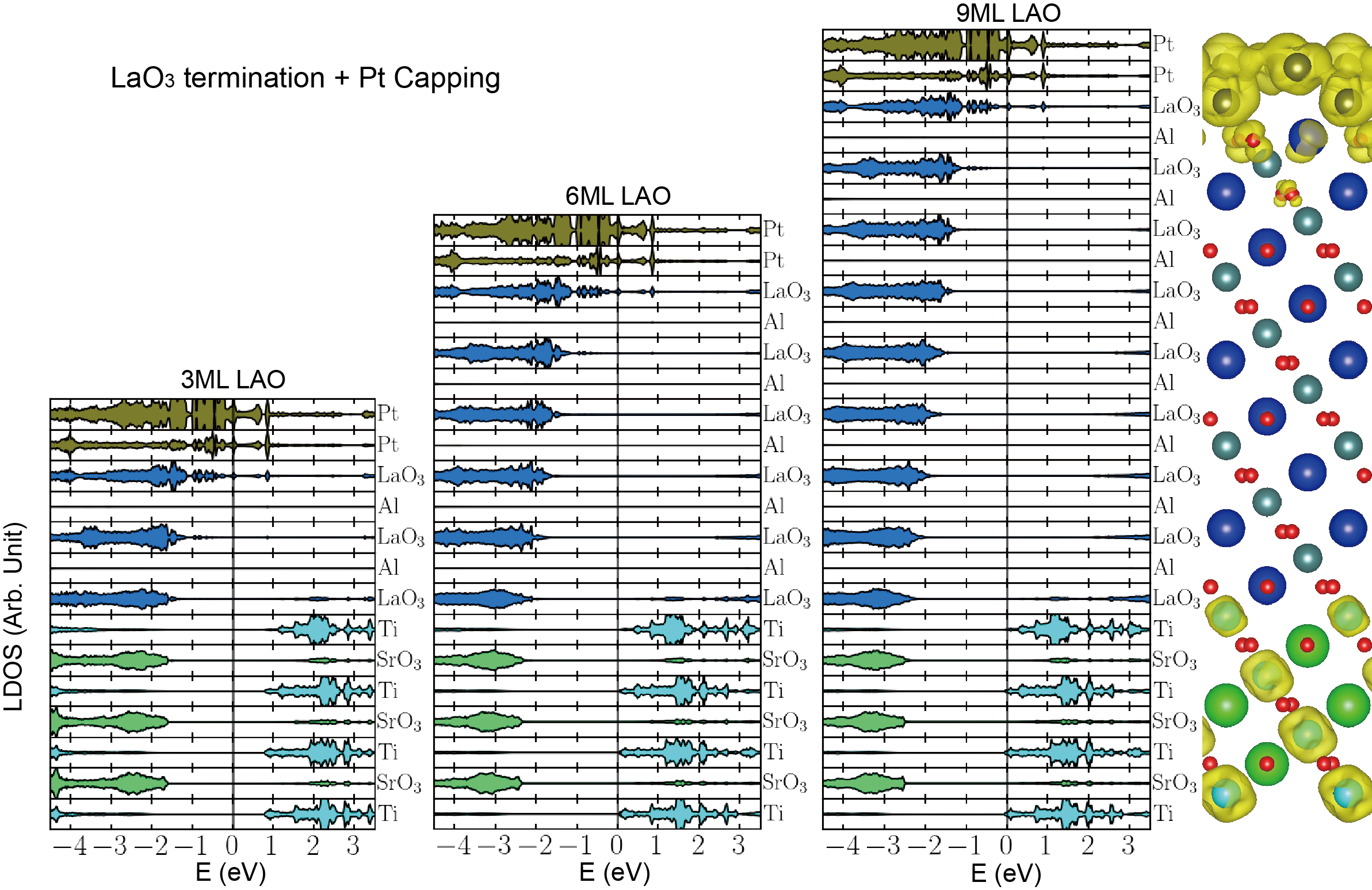}
\caption{Layer-resolved density of states (LDOS) of (111)\STO/\LAO~ slabs containing 7 ML of \STO~ and $N$ ML of \LAO~ capped with 1 ML of Pt ($N=3$, 6, 9). A side view of half of the simulation cell and the electron density distribution integrated from -0.2 to the Fermi level for 9 ML \LAO~ with 1ML Pt on top is shown in the far right panel.}
\label{LDOS2}
\end{figure*}

\begin{figure*}[htbp]
\centering
\includegraphics[width=0.45\textwidth]{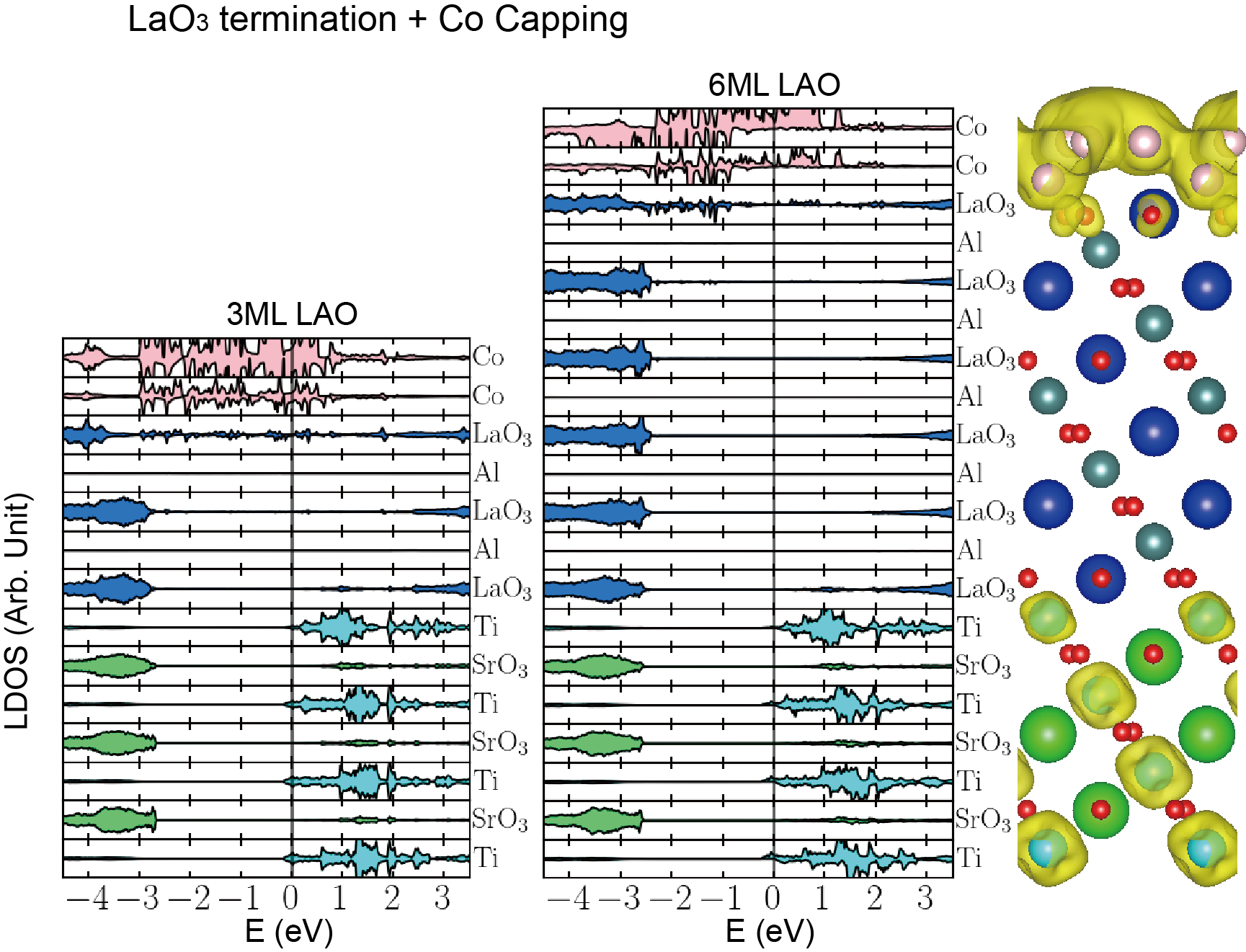}
\caption{The LDOS of (111)\STO/\LAO~ slabs containing 7 monolayers of \STO~ and three and six monolayers of \LAO~ capped with 1 ML of Co. A side view of half of the simulation cell and the electron density distribution integrated from -0.2 to the Fermi level for 6 ML \LAO~ with 1 ML Co on top is shown in the right panel.}
\label{LDOS3}
\end{figure*}
The experimental results presented above indicate that for (111)\STO/\LAO/metal $t_c$ is reduced. To assess the role of the metal capping, we covered the \LAO~ films with thicknesses of 3, 6, and 9 ML with Pt and Co. The calculated LDOS for a single monolayer of Pt and Co capping layers are displayed in Fig. \ref{LDOS2} and \ref{LDOS3}, respectively. As can be seen, for the Pt capping, the Fermi level touches the conduction band at the LaO$_3$/Ti interface when the \LAO~ thickness reaches 6 ML, whereas for the Co capping, the insulator-metal transition emerges already at a \LAO~ thickness of 3 ML, consistent with the experiment. The 6ML \LAO~ with Co capping represents the only case where a notable spin polarisation of the 2DES is observed with magnetic moments of Ti ranging from 0.11-0.20 $\mu_{\rm B}$.  To understand the observed critical thickness with metal capping, we list the calculated work functions of \LAO~ thin films on (111)\STO~ for different \LAO~ thicknesses, surface terminations, surface hydrogenation, and metal capping in Table. \ref{tab1}. For comparison, the work function of free-standing Pt(111) and Co(111) slabs (7 ML thick) unstrained and strained at the lateral lattice constant of \STO~ are also listed in Table. \ref{tab1}. We observe that the work function is weakly influenced by the \LAO~ thickness and is $\sim$5.0-5.2 eV for the LaO$_3$ termination and is reduced to $\sim$4.2-4.6 eV for the LaO$_2$(OH) termination. While the work function for the Co capping is $\sim$4.5-4.9 eV, it is significantly higher for the Pt capping ($\sim$ 6.1 eV). This is consistent with the higher critical thickness of the latter capping.  Nevertheless, the work function alone cannot account for the reduction of the critical thickness with respect to the uncovered \LAO~ film. On the other hand, a distinct behavior is observed from the layer-resolved DOS in  \ref{LDOS2} and \ref{LDOS3} for the systems with Co and Pt capping. While for Co capping the internal electric field within \LAO~ is quenched, for Pt there is still a considerable internal electric field, albeit smaller than for the bare \LAO~ film. Thus the different critical thicknesses for Co and Pt can be rationalized by the the different size of $p$-type Schottky barriers that form between \LAO~ and the metal contact: 2.5 eV (Co) vs. 1 eV (Pt). In contrast for the (001)-oriented interface the $p$-type Schottky barriers were similar for a Co and Pt contact ($\sim 2.3$ eV)~\cite{MetalcapDFT}, leading to similar band diagrams despite the difference in work function. 
\begin{table}[hb]
\centering
\caption{Calculated work functions (in eV) for different \LAO~ thicknesses (3, 6, and 9 ML), surface terminations (Al and LaO$_3$), partial surface hydrogenation [LaO$_2$(OH)], different metal cappings (Pt and Co), and of free-standing Pt(111) and Co(111) slabs of 7 ML thickness unstrained and strained to the lateral lattice constant of STO.}\label{tab1}
\begin{tabular}{cccccc}
  & 3 ML & 6 ML & 9 ML  \\
\hline
Al  & 4.478  &  4.599   & 4.534  \\
LaO$_3$  & 5.153  &  5.042   & 5.146  \\
LaO$_2$(OH)   & 4.186 &  4.193  & 4.631 \\
LaO$_3$/Pt   & 6.080  &  6.076   & 6.072  \\
LaO$_3$/Co   & 4.876 &  4.487  &  \\
\hline\hline
 & Pt(111) & Co(111) &  \\
\hline
Free-standing  & 5.688  &  5.121 & \\
Strained@$a_{\textrm{STO}}$  & 5.734  &  4.780 & \\
\end{tabular}
\end{table}
The right panels in Figs. \ref{LDOS1}, \ref{LDOS2} and \ref{LDOS3} also show the distribution of the quasi 2DES within (111)\STO~ for (111)\LAO/\STO~ with and without a metal capping layer. In all cases, a predominant $e_g^\prime$ orbital polarization is observed. We also note that for the systems covered by Pt or Co, a second conducting channel is present in the surface layer.
 
\par
\subsection*{Effect of metal capping on superconducting properties}
\begin{figure}[h]
\includegraphics[width= 8.3 cm]{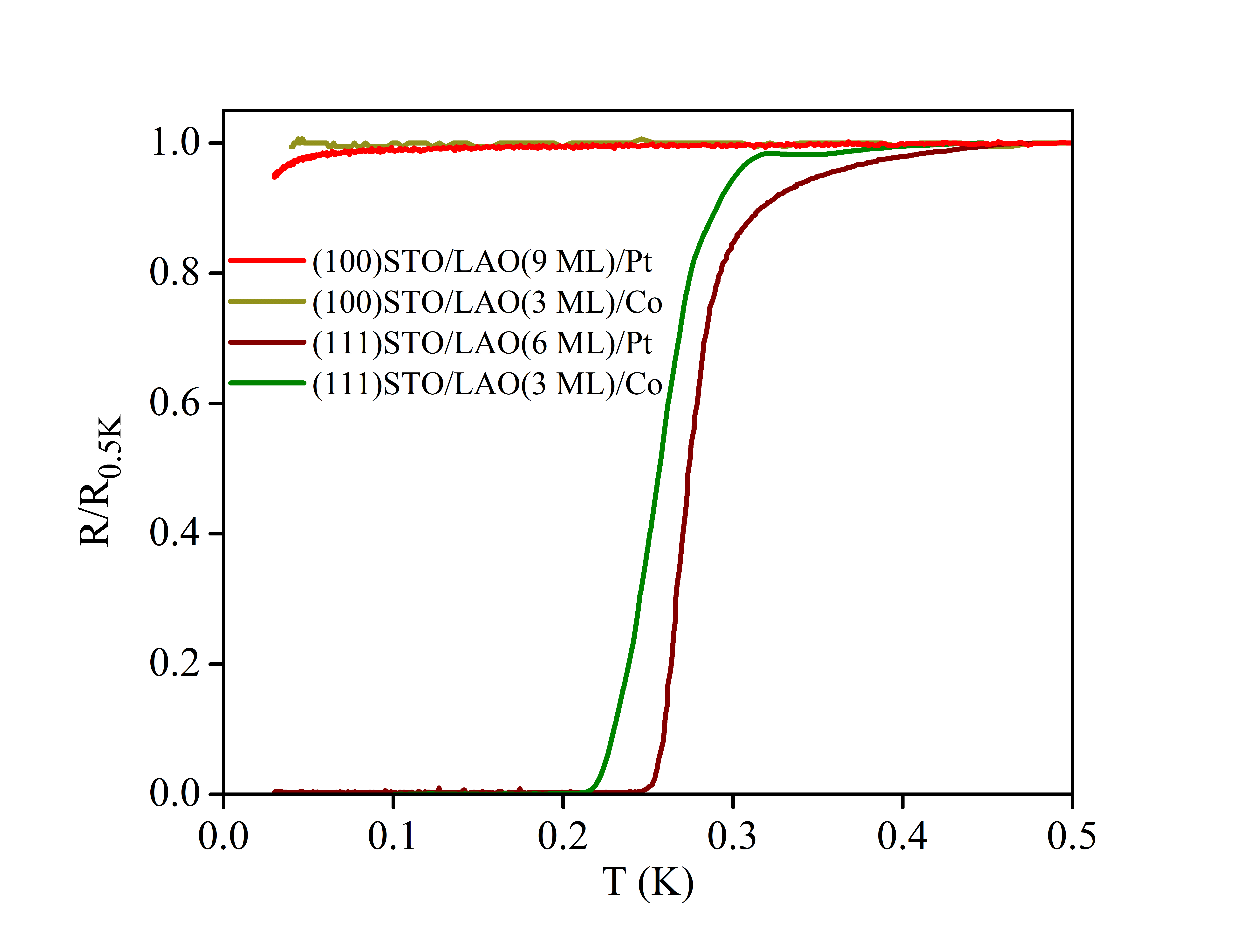}
\centering
\caption{The normalized resistance (R/R$_{(0.5 K)}$) of (100) \STO/\LAO(3ML)/Co/AlO$_x$, (100)\STO/\LAO(9ML)/Pt, (111) \STO/\LAO(3ML)/ Co/AlO$_x$, and (111) \STO/\LAO(6ML)/Pt. The \LAO/\STO (111) interfaces capped with Co and Pt show a superconducting transition and the corresponding critical thickness of \LAO~ is 3 and 6 Monolayer respectively.\label{RT}}
\end{figure}

The DFT+$U$ calculations presented above unveil the origin of the critical thickness for a metal-to-insulator transition in (111)\STO/\LAO~ and the role of metal capping. 
We now turn to study the effect of metal capping on superconductivity for the (100) and (111) interfaces. 
Surprisingly, we find that all the (111) samples, which show conductivity upon metal capping, also show superconductivity. In table 2 and S1 (Supplemental information), we summarize the properties of the (100) and (111) interfaces with various \LAO~ thickness and different metal capping. Fig.\ref{RT} displays the normalized resistance (R/R$_{(0.5 K)}$) as a function of temperature for different (100) and (111) interfaces at the critical thickness $t_c$ with Co and Pt capping. 
\begin{table*}[htb]
\caption{ Summary of the different samples measured for various thickness of \LAO~ upon different metal capping and corresponding nature of the interface. Note: For (100) interface one monolayer corresponds to one unit-cell (see also Table S1 (Supplemental information) for more samples).}
\centering
\begin{tabular}{|c|c|c|c|c|}
 \hline
 {Interface}
 & \multicolumn{1}{|p{1cm}|}{\centering {Metal} \\ {layer}}
 & \multicolumn{1}{|p{1.7cm}|}{\centering {Thickness} \\ {(ML)} }
 & {Conducting}
 & \multicolumn{1}{|p{2cm}|}{\centering {Super-} \\ {Conducting} }
 \\
 \hline
  (111) & Co & 2 & No & No\\
 \hline
 (111) & Co & 3  & Yes & Yes\\
 \hline
 (100) & Co & 2  & Yes & No\\
 \hline
 (100) & Co & 3  & Yes & No\\
 \hline
 (111) & Pt & 5 & No & No\\
 \hline
 (111) & Pt & 6  & Yes & Yes\\
 \hline
 (100) & Pt & 9  & Yes & No\\
 \hline
 (100) & Ag & 3  & Yes & No\\
 \hline
 \end{tabular}
\label{Table 1}
\end{table*}
\begin{figure}[h]
\centering
\includegraphics[width= 8.3 cm]{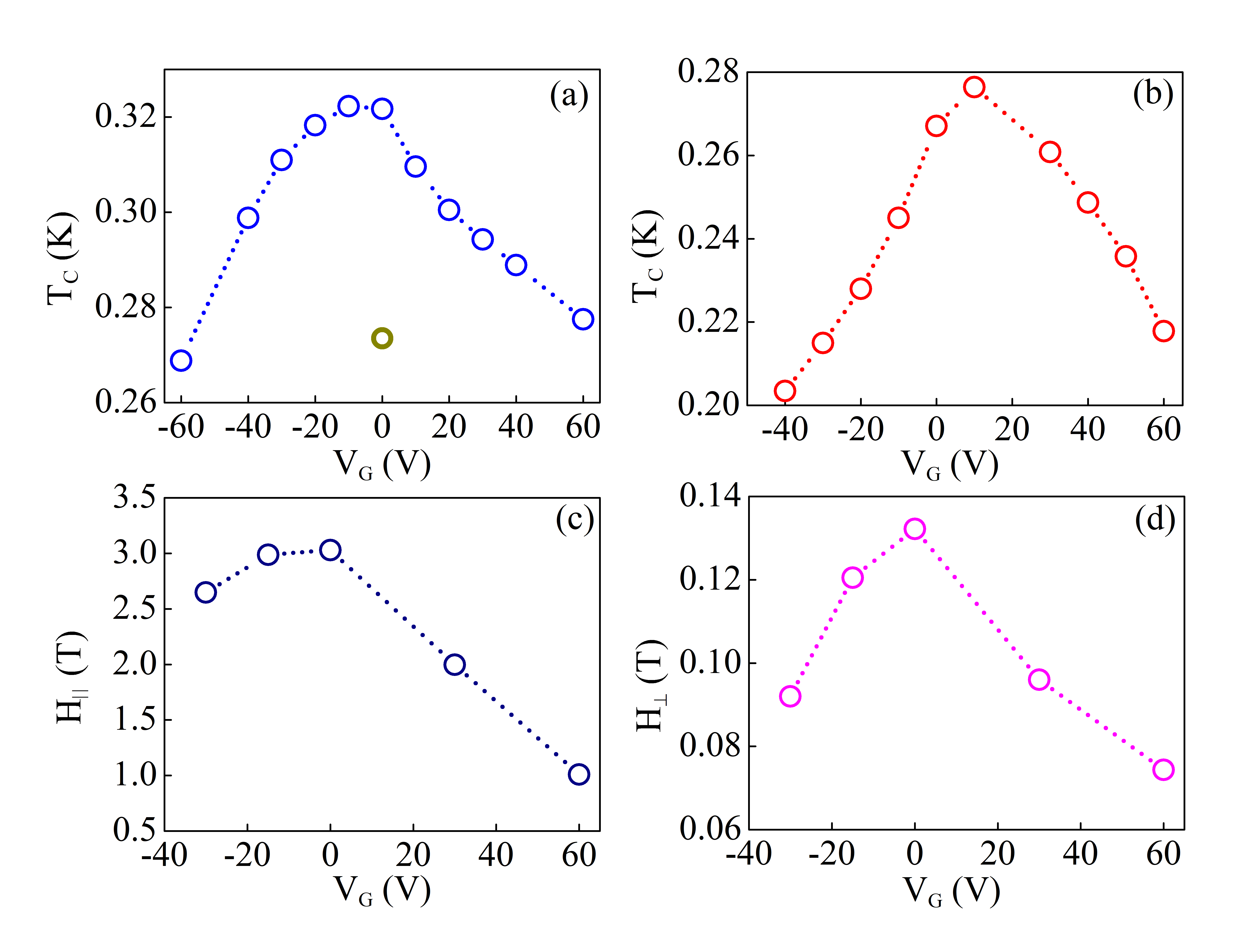}
\caption{(a) ,The  critical temperature (T$_c$) of (111) \STO/\LAO(6ML)/Pt  interface, the dark yellow circle shows the T$_c$ of the as cooled film. (b), (c) and (d) are superconducting critical temperature, perpendicular critical field and parallel critical field respectively of the (111) \STO/\LAO(3ML)/Co/AlO$_x$ interface as a function of gate voltage. The values of these critical parameters are in good agreement with that of the uncapped interface. Note:T$_c$ is defined as a temperature where the value of resistance drops  by $50\%$ of it's value at 0.5 K.\label{critical parameters} }
\end{figure}
\par
To make sure that the observed superconductivity is a two-dimensional (2D) interfacial effect and not a spurious one resulting from the metal deposition process, we studied the temperature dependence of the perpendicular and parallel critical field. We show in the supplementary information that they both follow the expected 2D Ginzburg-Landau temperature dependence (See Figure S4 of supplemental information).
\par
In Fig.\ref{critical parameters} we show the behavior of the critical temperature and critical fields as a function of gate voltage for Co and Pt capped (111) interface. The dome-shaped gate dependence and the values of $T_c$, the perpendicular critical field $H_\perp$, and the parallel critical field $H_\parallel$ are similar to the bare (111) \STO/\LAO~ interface \cite{khanna2019symmetry,Rout111}.
\par
One may claim that the absence of superconductivity in the Co-capped (100) \STO/\LAO~ is due to the Co ferromagnetism. Indeed, as shown previously in Fig.\ref{Hall}, the Co layer shows an anomalous hall effect. However, for such a thin cobalt film, one expects the magnetic coupling to be short-ranged with a negligible effect on the 2DES. Nevertheless, to eliminate the possibility that the close proximity ($\sim$12\AA) of the ferromagnetic Co to the interface is responsible for the absence of superconductivity in the (100) \STO/\LAO/Co/AlO$_x$, we measured (non-magnetic) Ag capped (100) interface with 3 ML of \LAO, which also shows no superconductivity, further demonstrating that the absence of superconductivity in capped (100) interfaces is not related to the ferromagnetism in the metal cap. In Figure S5 (Supplemental information), we show the temperature and gate bias dependence of the resistance for Ag capped (100) interface.
\par

How can we understand the robustness of superconductivity in (111) interfaces? Previous experimental studies on (111) \STO/\LAO~ interfaces show that even at strong negative gate voltages, superconductivity remains intact \cite{Rout111, morgbiSIT}. On the theory side, the curvature of the Fermi contour changes quickly upon charge accumulation, and both the conducting and superconducting bands get an equal contribution from the three degenerate t$_{2g}$ orbitals\cite{khanna2019symmetry}. By contrast, for (100) interface, there are distinct less mobile, non-superconducting band and a mobile superconducting band due to the very different effective masses of the light $d_{xy}$ and heavy $d_{yz}$, $d_{xz}$ bands. It is possible that the inter-band repulsion \cite{maniv100} results in shifting the second band to higher energy leaving only the metallic state.
Our DFT+$U$ calculations show that the e$_{g}^{\prime}$ bands are always present for the (111) conducting interfaces suggesting that this type of band plays a significant role in the observed superconductivity.

\section*{Methods}
Epitaxial \LAO~ films with different thicknesses were grown on Ti, and TiO$_2$ terminated atomically smooth (111), and (100) \STO~ substrates respectively at an oxygen pressure of \(1 \times 10 ^{-4}\) Torr and temperature 780$^{\circ}$C using pulsed laser deposition. The thickness was in-situ monitored by reflection high energy electron diffraction (RHEED) (see supplementary information Figure S1). The samples were then transferred to a metal deposition chamber (e-beam for the Co and Ag and Sputtering or e-beam for the Pt) where they were pre-annealed for two minutes at 200$^{\circ}$C and at a pressure of \(1 \times 10 ^{-8}\) Torr to remove surface contaminants. Metallic layers of $\approx$3 nm of platinum (Pt), silver (Ag), or cobalt (Co) were deposited at room temperature. For the latter, we used an additional 3 nm AlO$_x$ to prevent oxidation. Wire bonding was used to connect to the sample electrically. 
Density functional theory (DFT) calculations were performed on thin \LAO~ films on (111)-oriented \STO, using the projector augmented wave (PAW) method \cite{PAW} as implemented in the VASP code \cite{VASP}. The generalized gradient approximation was used for the exchange-correlation functional, as parametrized by Perdew, Burke, and Ernzerhof \cite{PBE}. Static correlation effects were considered within the DFT$+U$ formalism \cite{LDAU2}, employing $U=3$ eV for the Ti $3d$ orbitals, in line with previous work \cite{HubbardU1,HubbardU2,HubbardU3}. The \LAO~ thin films on (111)\STO~ were modeled in the slab geometry with two symmetric surfaces to eliminate spurious electric fields. The (111)\STO/\LAO~ slabs contain 7 monolayers (ML) of \STO~ (substrate) and 3-9 ML of \LAO~ on both sides of the substrate. Additionally, in order to assess the role of metallic contacts, a Pt and Co ML was added on top of the \LAO~ film. The modeled slabs contain $\sim$60 atoms for 3 ML \LAO~ and $\sim$120 atoms for 9 ML \LAO, depending on surface termination and metal capping. The lateral lattice constant of the modeled (111)\STO/\LAO~ slabs was fixed to the \STO~ substrate ($\sqrt{2}a\times\sqrt{2}a$ lateral unit cell with $a=3.905$ {\AA}). A vacuum region of 15 {\AA} was adopted to minimize the interaction between the slab and its periodic images. A cutoff energy of 600 eV was used to truncate the plane-wave expansion and a $\Gamma$-centered $k$-point mesh of $12\times12\times1$ to sample the Brillouin zone (BZ). The atomic positions were fully optimized taking into account octahedral rotations and distortions until the forces on all atoms were less than 0.01 eV/{\AA} and the change in total energy was less than $10^{-6}$ eV. Spin polarization was also considered in the DFT+$U$ calculations to account for possible magnetic moments of the Ti $3d$ electrons and the metal capping layer.

\bibliography{arxiv}

\section*{Acknowledgements}

We thank Moshe Goldstein and Udit Khanna for useful discussions. This research was supported by the Bi-national Science Foundation under grant 2014047 and by the Israel Science Foundation under grant : 382/17. RP and LL acknowledge computational time at the Leibniz Rechenzentrum, project pr87ro.

\section*{Author contributions statement}

Y.D., A.G.S., and H.Y.H conceived the experiment. R.S.B., M.M.,and P.K.R. made the samples. H.Y, A.G.S. and R.S.B. did the metallization. R.S.B., M.M., P.K.R., and G.T. conducted the transport measurements and analyzed the data. R. P. conceived the theoretical modelling and L.L.L. performed the DFT calculations. R.S.B., L.L.L, R.P., and Y.D. wrote the paper. All authors reviewed the manuscript. \\

\newpage
\textbf{Supplementary information for Concomitant appearance of conductivity and superconductivity in (111)  LaAlO$_3$/SrTiO$_3$ interface with metal capping}\\

\par
\textbf{S1: Reflection-high energy electron diffraction (RHEED) oscillations for \LAO~ film grown on \STO~(111) and (100) substrates respectively}\\
\begin{figure}[h]
\includegraphics[width=3.5 in]{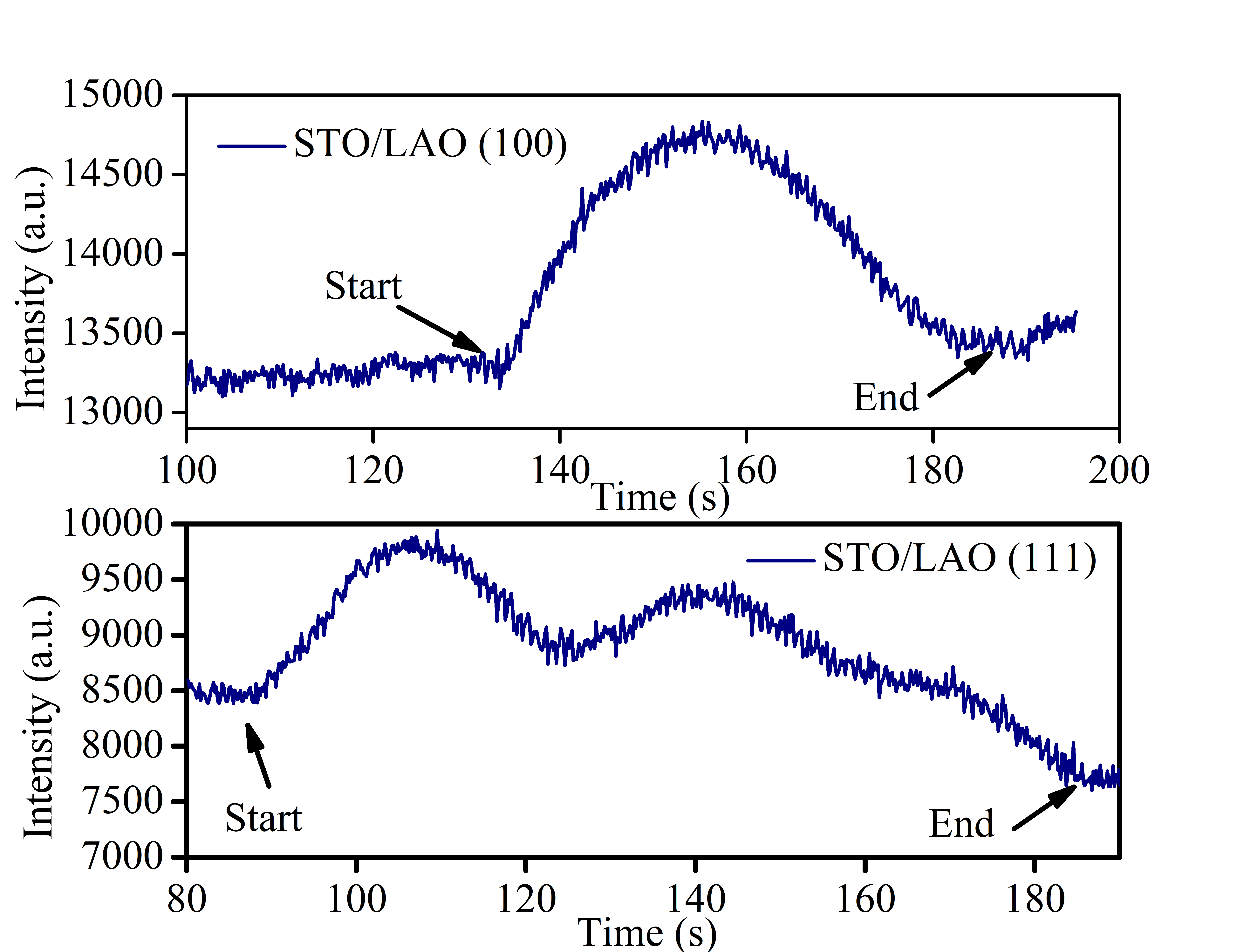}
\centering
\caption{S1 : Reflection-high energy electron diffraction (RHEED) oscillations monitored during the deposition of \LAO~ on \STO~(111) and (100) substrate suggests a layer by layer growth of the film. Each oscillation corresponds to the one monolayer.}
\end{figure}\\
\par
\textbf{S2: Normalized resistance (R/R$_{(-400 V)}$) of (111) \STO/\LAO(6ML)/Pt }\\
\begin{figure}[h]
\includegraphics[width=3.5 in]{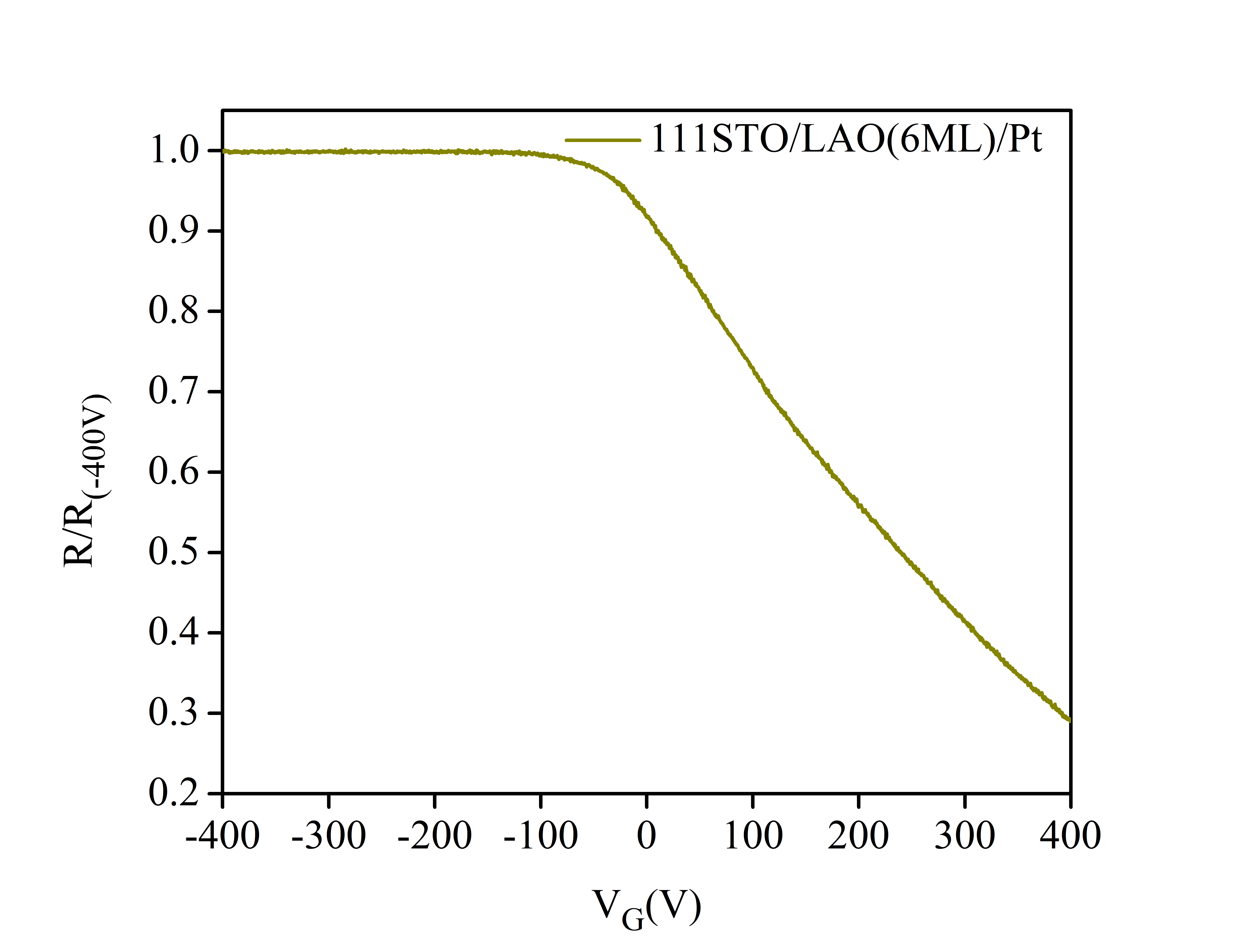}
\centering
\caption{S2 : The normalized resistance (R/R$_{(-400 V)}$) of (111) \STO/\LAO(6ML)/Pt sample suggests a complete depletion of 2DES at negative gate voltages and a dominating contribution of metal overlayer.}
\end{figure}\\

\textbf{S3: Sheet Resistance of two dimensional electron system without Capping}\\
The sheet resistance of the two dimensional electron system (2DES) without capping was extracted from the analysis of the parallel resistance model as given in the equation 1. Where $R_{Measured}$, $R_{Metal- Overlayer}$,and $R_{2DES}$ is the experimentally measured resistance of the parallel conducting channel, resistance of metal overlayer, and resistance of the 2DES respectively. In our analysis we assumed $R_{metal-capping}$= $R_{Measured}$ at strong negative gate voltages. Figure S3 represents the extracted sheet resistance of the 2DES for (111)\STO/\LAO/Co/AlO$_x$ and (111)\STO/\LAO/Pt samples for 3 and 6 ML \LAO~ as a function of gate voltages. The diverging nature of the resistance at negative gate voltage is expected for a depleted 2DES.
\begin{equation}
    \frac{1}{R_{Measured}}= \frac{1}{R_{Metal- Overlayer}}+\frac{1}{R_{2DES}}
    \end{equation}
\par
\begin{figure}[h]
\includegraphics[width=3.4in]{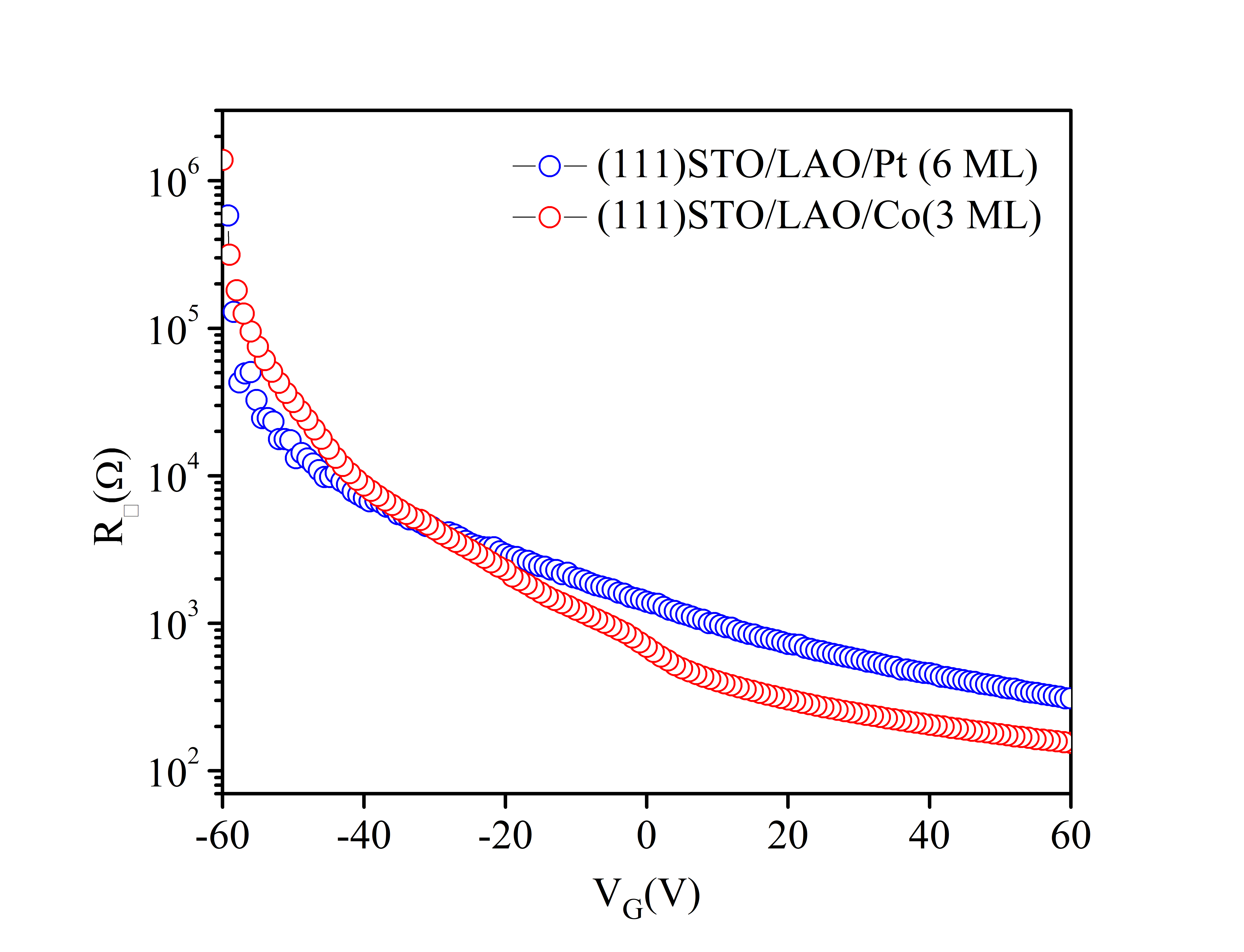}
\centering
\caption{S3 : The extracted sheet resistance of (111)\STO/\LAO/Co/AlO$_x$ and (111)\STO/\LAO/Pt samples for 3 and 6 ML of \LAO~ as a function of a  gate voltages. The resistance was extracted according to the equation 1. A diverging nature of the resistance at negative gate represents the depletion of the 2DES.}
\end{figure}
\par
{\textbf{Table S1:} The extension of table 1 from the main text. Summary of the different samples measured for various thickness of LaAlO$_3$  upon different metal capping and corresponding nature of the interface.}
\begin{table}[h]
\begin{tabular}{|c|c|c|c|c|} 
 \hline
 \textbf{Interface}
 & \multicolumn{1}{|p{1cm}|}{\centering \textbf{Metal} \\ \textbf{layer}}
 & \multicolumn{1}{|p{1.7cm}|}{\centering \textbf{Thickness} \textbf{(ML)}\\  } 
 & \textbf{Conducting} 
 & \multicolumn{1}{|p{2cm}|}{\centering \textbf{Super-} \\ \textbf{Conducting} }\\ 
 \hline
  (111) & Co & 1  & No & No\\ 
 \hline
  (111) & Co & 2 & No & No\\ 
 \hline
 (111) & Co & 3  & Yes & Yes\\
\hline
 (111) & Co & 4 & Yes & Yes\\ 
 \hline
 (111) & Co & 6  & Yes & Yes\\ 
 \hline
 (111) & Co & 7  & Yes & Yes\\ 
 \hline
 (111) & Pt & 5 & No & No\\
 \hline
 (111) & Pt & 6  & Yes & Yes\\
 \hline
 (111) & Pt & 9  & Yes & Yes\\
 \hline
 (111) & Pt & 12  & Yes & Yes\\
 \hline
 (100) & Co & 2  & Yes & No\\
 \hline
 (100) & Co & 3  & Yes & No\\
 \hline
 (100) & Pt & 9  & Yes & No\\
 \hline
 (100) & Ag & 3  & Yes & No\\  
 \hline
\end{tabular}
\end{table}
\par
\textbf{S4: Analysis of critical fields}\\
Temperature dependence of perpendicular and parallel critical fields were analyzed according to the phenomenological Ginzburg-landau theory for the 2-dimensional nature of the superconductivity. The perpendicular and parallel critical fields in the framework of the Ginzburg-landau theory follow equation 2 and 3 respectively. In Figure S4, we show the fit to equation 2 and 3 for (111)SrTiO$_3$/LaAlO$_3$(3ML)/Co interface. The extracted perpendicular ($H_\perp$) and parallel critical field ($H_\parallel$) as a fuction of gate voltages has been shown in Figure 7 of the main text.
 \begin{equation}
    H_\perp= \frac{\phi_0}{2\pi\xi^2} (1-\frac{T}{T_c})
    \end{equation}
\begin{equation}
    H_\parallel= \frac{\phi_0\sqrt12}{2\pi\xi^2d} (1-\frac{T}{T_c})^\frac{1}{2}
    \end{equation}
\begin{figure}[h]
\centering
\includegraphics[width=4.5 in]{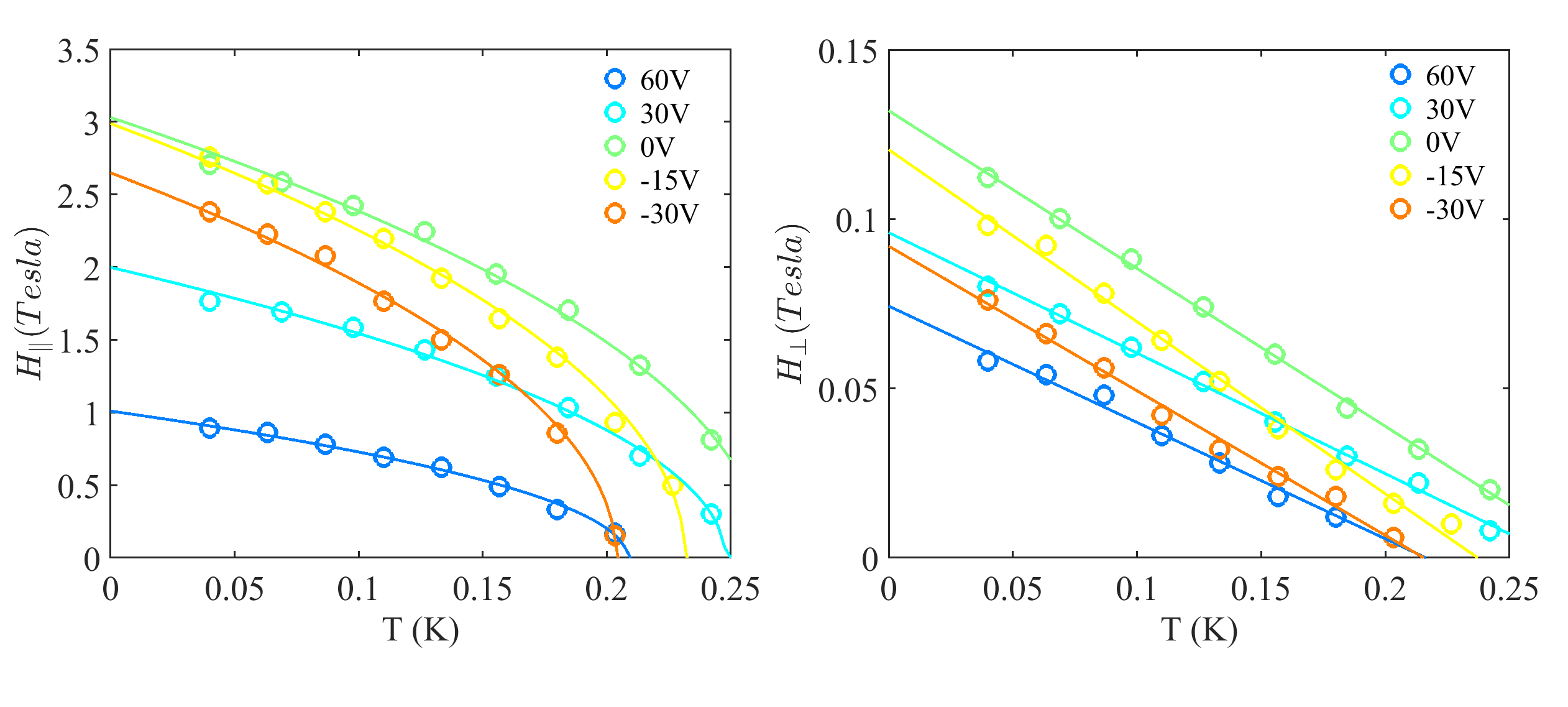}
\caption{S4: The perpendicular ($H_\perp$) and parallel critical field ($H_\parallel$) as a fuction of temperature for (111)SrTiO$_3$/ LaAlO$_3$(3ML)/Co interface. The solid lines are fit to the equation 2 and 3. }  
\end{figure}
\par
\textbf{S5: Role of Ag Capping}\\
To show that absence of superconductivity in co-capped (100) \STO/\LAO~ interface is not related to ferromagnetism in the cobalt, we used silver (Ag) capping. Ag has low work function and can therefore reduce the critical thickness by increasing the charge transfer. The measured Ag capped (100) interface for 3 monolayers (ML) of LaAlO$_3$ does not show full superconducting transition for any gate voltage. This confirms that ferromagnetism of Co is not the cause for the absence of superconductivty in (100) co-capped interfaces.  In Figure S5 we show the sheet resistance of Ag capped (100)\STO/\LAO~ as a function of temperature at different positive gate voltages.  A dip in the sheet resistance can be attributed due to the localized superconducting regions which are not connected percolatively even on application of positive gate bias. These localized regions may arise from a slight thickness nonuniformity. The data thus suggest there is no macroscopic superconductivity present in (100)\STO/\LAO~ interface below the bare critical thickness upon metal capping.
\begin{figure}[h]
\centering
\includegraphics[width=3 in]{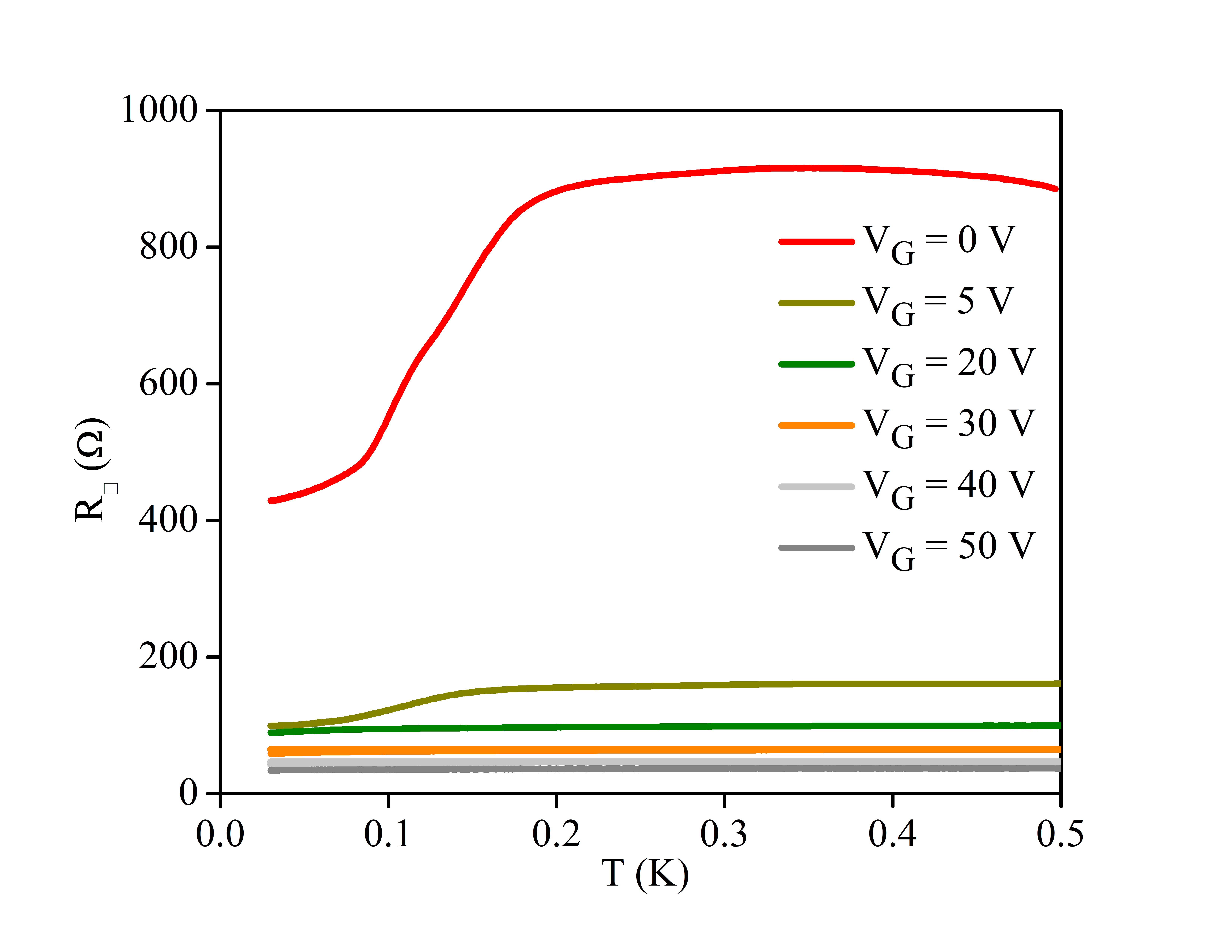}
\caption{S5: The sheet resistance of (100) SrTiO$_3$/ LaAlO$_3$/Ag interface for 3 \LAO~ ML as a function of temperature at different gate voltages shows an absence of  superconductivity. A dip in the resistance can be attributed to the localized superconducting regions which may result from a thickness inhomogenity of \LAO~.}
\end{figure}\\

\end{document}